# A bibliometric view of AI Ethics development

Di Kevin Gao
Management Department - California State University, East Bay
Hayward, CA, USA
kevin.gao@csueastbay.edu

Andrew Haverly
Computer Science and Engineering Dept
Mississippi State University
Mississippi State, MS 39762
arh876@msstate.edu

Dr. Sudip Mittal
Computer Science and Engineering Dept
Mississippi State University
Mississippi State, MS 39762
mittal@cse.msstate.edu

Dr. Jingdao Chen
Computer Science and Engineering Dept
Mississippi State University
Mississippi State, MS 39762
chenjingdao@cse.msstate.edu

***Abstract*—** Artificial Intelligence (AI) Ethics is a nascent yet critical research field. Recent developments in generative AI and foundational models necessitate a renewed look at the problem of AI Ethics. In this study, we perform a bibliometric analysis of AI Ethics literature for the last 20 years based on keyword search. Our study reveals a three-phase development in AI Ethics, namely an incubation phase, making AI human-like machines phase, and making AI human-centric machines phase. We conjecture that the next phase of AI ethics is likely to focus on making AI more machine-like as AI matches or surpasses humans intellectually, a term we coin as "machine-like human".

## I. Introduction

Artificial Intelligence (AI) Ethics is the study of the ethical and responsible development and deployment of AI technology. Our bibliometric analysis of AI Ethics literature published between 2004 and 2023 points us to a three-phase development: 1. Incubation; 2. Making AI human-like machines; 3. Making AI human-centric machines.

This article contributes to AI Ethics discussions with unique insights based on keyword usage patterns. It also contrasts "human-like machine", "human-centric machine", and "machine-like human", which represent the past, current, and potential future phases of AI Ethics development.

## II. Definitions and historical development

AI was first coined in 1955 at the Dartmouth Workshop [1]. John McCarthy, one of the key contributors at the conference and an AI pioneer, defined AI as "the science and engineering of making intelligent machines, especially intelligent computer programs. It is related to the similar task of using computers to understand human intelligence, but AI does not have to confine itself to methods that are biologically observable" [2].

AI has gone through multiple cycles of boom and bust. It was a rising star from its inception to 1973. Scientists were excited by its potential to solve algebra word problems, prove geometry theorems, and even learn to speak. However, it failed to deliver on the hyped expectations. That led to the Lighthill Report in 1974, which triggered a massive loss of confidence in AI [3]. In the United States, the Defense Advanced Research Projects Agency (DARPA) also drastically reduced its AI funding. AI sank into an "AI Winter" until 1980. The relief in the 80s turned out to be short-lived. In 1987, AI was again placed in a freezer. By the early 2000s, AI was haunted by over-promises and under-delivery. In 2005, John Markoff in the New York Times described that some computer scientists avoided the term artificial intelligence altogether "for fear of being viewed as wild-eyed dreamers" [4]. In 2007, Alex Castro referred to artificial intelligence as a subject that has "too often failed to live up to their promises" [5]. That has resulted in "once something becomes useful enough and common enough it's not labeled AI anymore" [6]. In the 1990s and 2000s, new computer science disciplines flourished. However, they were deliberately not categorized under Artificial Intelligence, for example, informatics, machine learning, machine perception, analytics, predictive analytics, decision support systems, knowledge-based systems, business rules management, cognitive systems, intelligent systems, language models, intelligent agents, or computational intelligence.

AI regained its popularity with Google Translate, Google Image Search, and IBM Watson's winning the Jeopardy game in 2011 [7]. 2012 marked a turning point for AI due to breakthroughs in deep learning and GPU technology. AlexNet used GPU to train its Convolutional Neural Network (CNN) model to recognize and label images automatically and won the ImageNet 2012 Challenge by a large margin[8] [9]. AI's resurgence became insurmountable. AI broadened its scope to absorb many downstream research fields. It became an aggregator and a destination.

In November 2022, Open AI released ChatGPT which attracted intense interest from the general public. It triggered an all-out war between the Big Techs and stiffened competition between rival countries. Ethical AI development and deployment have become more important than ever.

## III. Methods

We selected SCOPUS[1] as the main data source and VOSviewer[2] as the data aggregator. In the SCOPUS database, we searched for "AI ethics" OR "artificial intelligence ethics" OR "machine ethics" OR "algorithm ethics" OR "information ethics" OR "ethics of technology" OR "Robotic Ethics" OR "Robot Ethics" OR "artificial moral agent" OR "artificial moral agents" from 2004 to 2023, a period of 20 years, for all languages, all countries, and territories. In total, 2,517 articles were selected. After removing 60 entries due to missing info, a total of 2,457 pieces of literature were included in this analysis. For keyword analysis, we used 2004-2023 co-occurrence author keywords from VOSviewer. We used "Full counting", which means each keyword is counted as one regardless of how many keywords were listed in the literature. We then exported the results for time series and pattern analysis.

## IV. AI Ethics Development

### A. AI ethics and related ethics usage analysis

We calculated the keyword usage frequencies for AI Ethics and other related ethical fields based on SCOPUS data. Other related ethical fields included information ethics, machine ethics, roboethics, technology ethics, computer ethics, data ethics, engineering ethics, digital ethics, and computational ethics. The data is summarized in Figure 1.

"AI Ethics" or "Artificial Intelligence ethics" first appeared in keywords in 2008, followed by five 5 years of hibernation. In 2014, AI Ethics reemerged and has since enjoyed exponential growth. In 2014, there was only one occurrence of the keyword "AI Ethics" in our literature research. However, by 2022, the keyword frequency had increased to 148. In the 2023 partial year till July 28, the usage has also reached 114. The keyword "AI Ethics" completely outnumbered the rest of the terms such as "roboethics", "data ethics", or "machine ethics".

This finding is important because in the historical development section of AI, we know that 2012 marked the turning point for AI as a research field. We believe 2014 is the year that AI Ethics was formed. Before that, AI Ethics was dispersed across information ethics, machine ethics, roboethics, technology ethics, and computer ethics. Thus, we define the pre-2014 period as the incubation period.

### B. Ethics principles usage pattern analysis

We leveraged VOSviewer to unpack keyword usages that are related to AI Ethics. Figure 2 is generated from the VOSviewer based on data since 2004 by using co-occurrence data and author keywords. We used full counting instead of partial counting, which means each author keyword is counted as one regardless of how many keywords were used in the literature. The size of the circle indicates the keywords' relative frequency. Color represents the closeness of the topics.

We sorted this information chronologically and further bifurcated the keywords based on product orientation and their associations with AI Ethics principles. The information is presented in Figure 3. The top rectangles are product-oriented features, the middle ovals illustrate when keywords started to be consistently used.

The results revealed a significant shift in AI Ethics research principles from 2020. Between 2014 and 2019, AI Ethics keywords focused on principles to make AI ethical humans, e.g., trust, empathy, justice, care, and fairness. From 2020, however, the AI Ethics keywords were increasingly focused on protecting the downside risks and making AI explainable, accountable, trustworthy, non-biased, non-discriminatory, less opaque, and for-diversity.

Based on this observation, we grouped AI Ethics development into the following three phases:

- Phase I: Incubation (2004 to 2013)
- Phase II: Making AI human-like machines (2014 to 2019)
- Phase III: Making AI human-centric machines (2020 and on)

### C. AI Ethics Development Phases

In the following section, we will go through each phase and highlight the major developments.

#### 1) Phase I: Incubation (2004-2013)

AI Ethics trailed AI's rapid development. In 1997, IBM's Deep Blue defeated the world chess champion Garry Kasparov [10]. In 2005, the Stanford autonomous vehicle, Stanley, successfully crossed 212 kilometers of terrain in the Mojave Desert [11]. In 2011, IBM's Watson won the Jeopardy! that conventional wisdom believed only humans could master [12]. AI technology's fast development galvanized many exciting research fronts and, in a way, "pushed" AI Ethics to the forefront. It became apparent that other ethics would not be sufficient to cover the wide spectrum of fields that AI covers. AI Ethics as a research field was born.

The popular keyword in 2004 was "privacy". Privacy was not a new discipline; neither was it exclusive to AI. It became a hot topic with the explosive growth of data collection and data usage in the Internet age. In the next few years, "autonomy", "reliability", "safety", "security", and "sustainability" surfaced. The majority of the keywords were product-oriented.

#### 2) Phase II: Make AI Human-like Machines (2014-2019)

In Phase II, AI has increasingly demonstrated its potential to function like a human. AI Ethicists and the general public welcomed the development. AI Ethics' focus was on the ethical application of AI, the mini-human, in different fields. For this reason, we labeled this phase "Make AI Human-like Machines".

During this phase, AI continued its breakneck advancement and pushed deep into new frontiers. In 2014, generative adversarial networks (GAN) were developed to synthesize new and creative images from existing ones. In 2015, AI enabled machines to "see" and label images better than humans [13]. In 2016, Deep Mind's AlphaGo defeated world Go champion Lee Sedol [14]. In 2018, AI beat human dermatologists in accurately detecting skin cancer [15]. In 2018, Google Waymo's Robotaxi started roaming in Phoenix's streets [16]. The general public viewed the development positively and was excited by AI's boundless applications.

[1] www.scopus.com

[2] Vosviewer.com

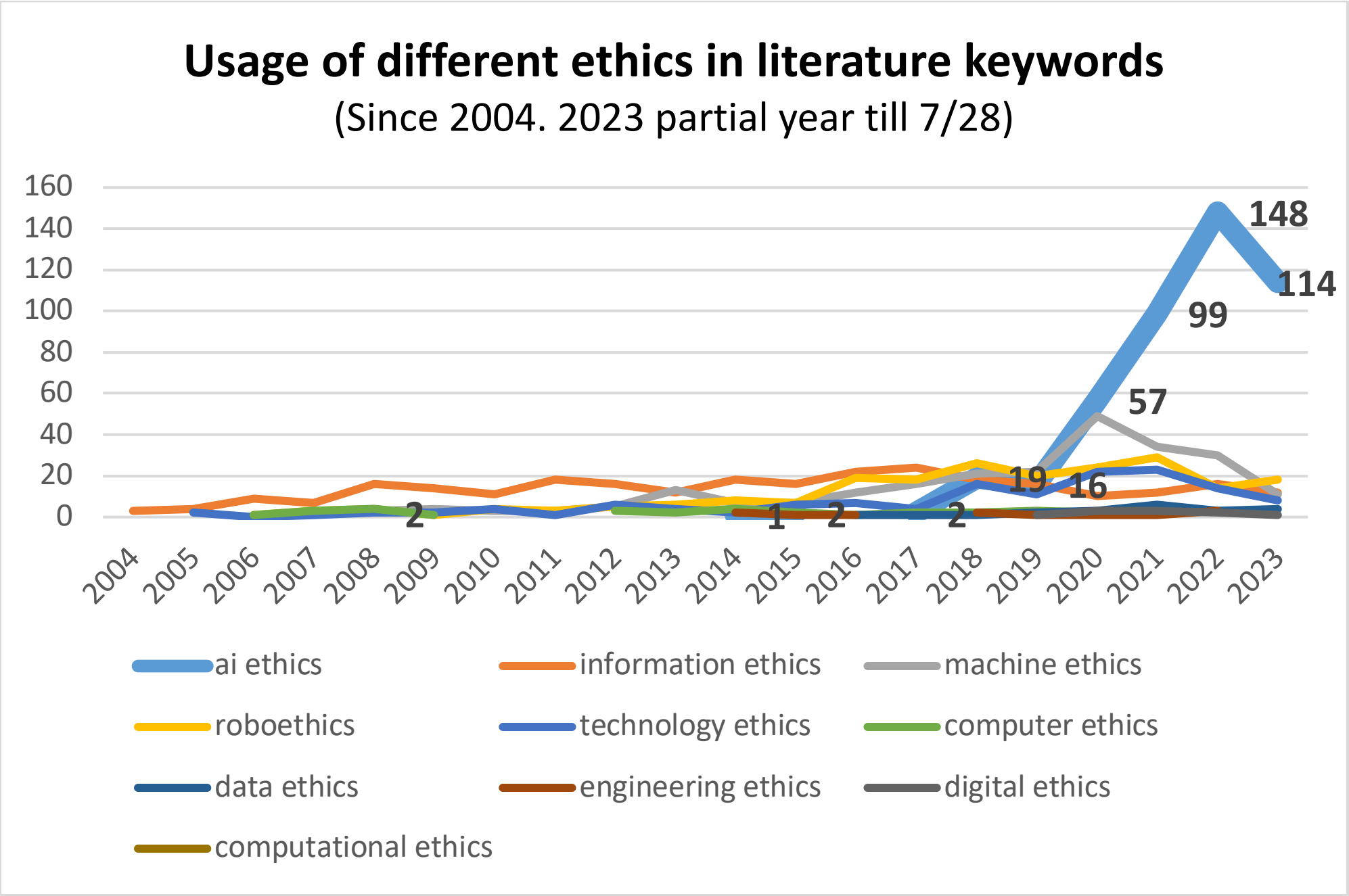

Fig 1: Usage of different ethics in literature key words. AI Ethics first appeared in 2008. It consistently appeared after 2014 and have taken off since 2020.

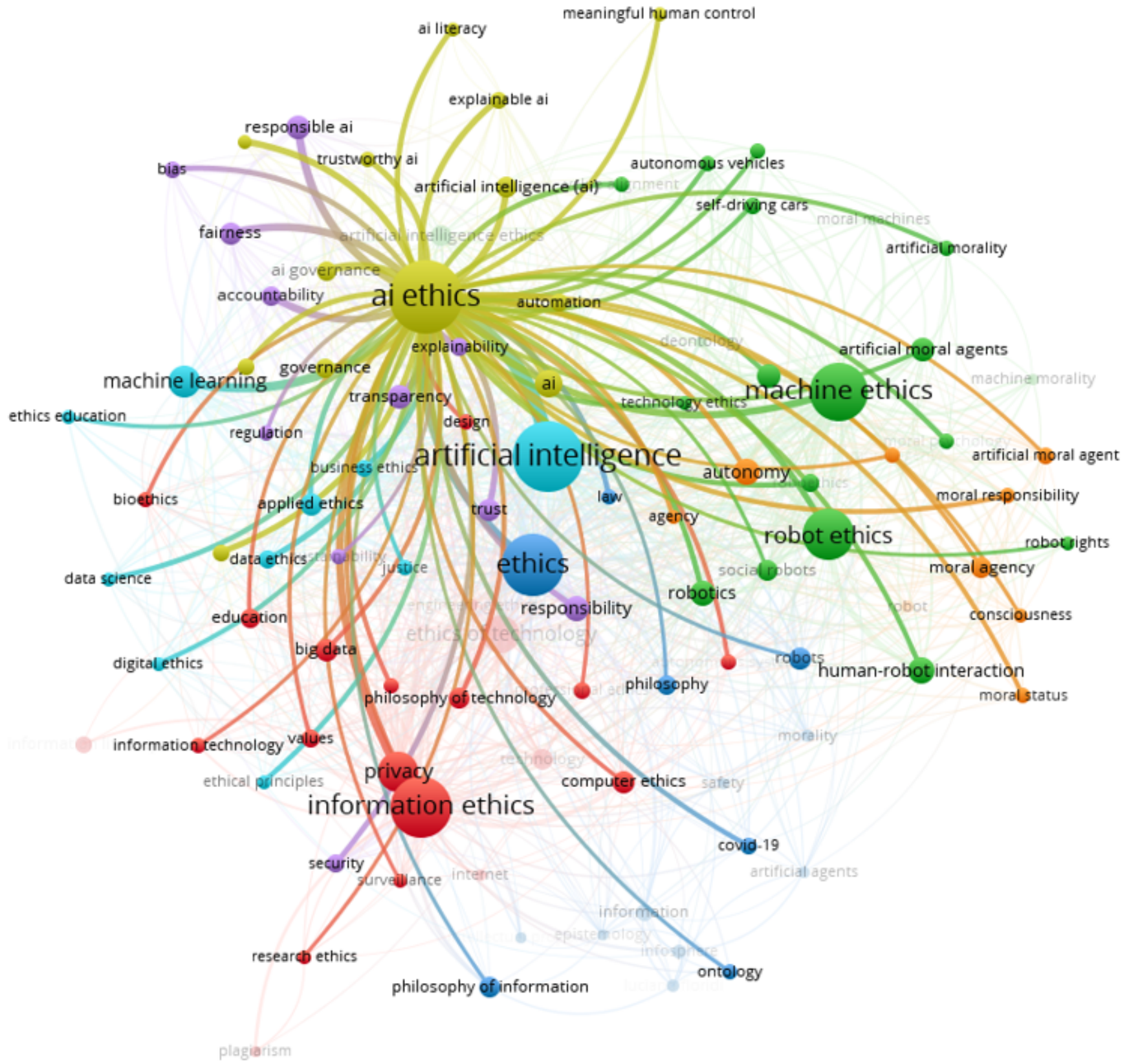

Fig 2: AI Ethics and related fields based on bibliographical data in VOSviewer by using Scopus data from 2004 to July 28, 2023.

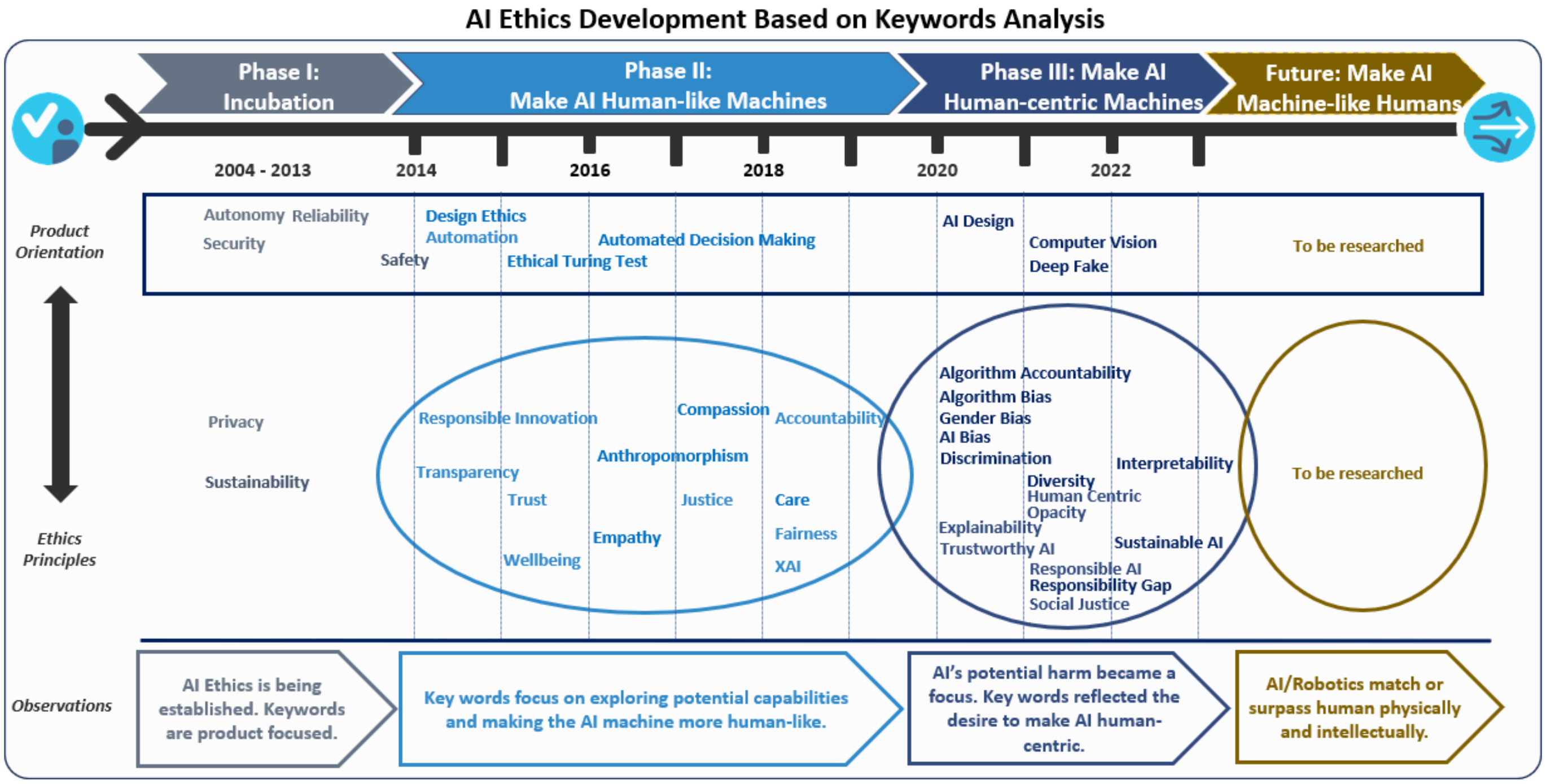

Fig 3: AI Ethics Development Phases Based on Keyword Analysis

However, AI Ethics development lagged behind AI technology development. As mentioned in the historical background section, AI regained popularity and broadened its scope to include any computer science disciplines that enabled human-like intelligence in 2012. Machine learning, machine perception, text analysis, natural language processing (NLP), logical reasoning, game-playing, decision support systems, data analytics, and predictive analytics became AI's upstream or supporting disciplines. Robotics (including autonomous vehicles) was a fast-developing field that was enabled by AI. AI became an aggregator.

AI Ethics keywords during this phase reflected many human-like features, for example, "accountability", "care", "compassion", "empathy", "fairness", "justice", "transparency", "trust", and Explainable AI (XAI). The AI Ethics community wanted to make this intelligent technology accountable, caring, compassionate, empathetic, fair, unbiased, transparent, and trustworthy, just like an ethical human.

*3) Phase III: Make AI Human-centric Machines (2020-present)*

In Phase III, while continuing its rapid ascension, AI had shown aspects that were far from angelic. The AI Ethics community focused on grounding AI into an explainable, responsible, and trustworthy machine that serves humans instead of being a runaway alien technology. That was the reason we titled this phase "Make AI Human-centric Machines".

By 2020, AI had surpassed humans in handwriting recognition, speech recognition, image recognition, read comprehension, and language understanding [17]. In the meantime, Deep Fakes exacerbated online misinformation and undermined basic human trust. In May 2021, the United States National Security Commission urged the US to win the AI arms race against China, reminiscing the costly and dangerous Cold War [18]. In July 2022, Google fired an engineer who claimed that its LaMDA language model was sentient, exacerbating the general public's suspicion of AI [19]. In November 2022, Open AI released Chat GPT 3.0 to the public and triggered an all-out AI race. It looked increasingly like the AI companies were racing to the bottom to win but put safety and security on the backburner [20]. In 2023, ChatGPT became the second Large Language Model to pass the Turing Test [21] [22]. In 2023, Goldman Sachs estimated that 300 million jobs could be displacement by AI [23]. Scientists, entrepreneurs, and public officers started to alarm the general public about the consequences of unconstrained AI development. Public distrust of AI surged.

During this phase, precautionary keywords showed up very frequently in the literature, such as "algorithm bias", "AI bias", "gender bias", and "discrimination" in 2020, and "opacity", "responsibility gap", and "social justice" in 2021. Meanwhile, keywords such as "explainability" and "trustworthy AI' popped up in 2020. "Human-centric", "responsible AI" showed up in 2021, and "interpretability" and "sustainable AI" showed up in 2022. The AI ethics community wanted to make AI responsible, explainable, and trustworthy to humans. AI Ethics entered a phase to make AI "human-centric".

### D. *The future of AI Ethics*

AI technology is disruptive in nature. AI Ethics is pivotal in the benign and benevolent rollout of AI technology. With the current development pace, it is almost inevitable that AI and robotics will match or surpass humans both physically and intellectually. AI can become near-human. AI ethicists may need to explore how to make these intelligent AI embodiments "machine-like humans": machines that are intelligent and capable, but never achieve the same status as full humans.

Another challenge is the confluence of AI, robotics, and biotechnology. Machine-augmented humans and machine-augmented non-human (animals or newly created species) could blur the definition of humans. AI ethicists may need to study the ethical ramifications of this development and draw redlines on socially acceptable creations of alien beings.

Superintelligence, a form of AI that is superior to human intelligence, can be a concern. Even if the risks are extremely low, if indeed it happens, the consequences could be incredibly serious. AI Ethicists may need to develop a framework to prevent Superintelligence from remotely happening.

## V. LIMITATION

Our bibliometric analysis has a few limitations:

1. We rely heavily on SCOPUS for data sources. Literature unlisted on SCOPUS would be excluded.
2. We relied heavily on VOSviewer to generate keyword co-occurrence data. Errors in VOSviewer could be carried into our final analysis.
3. Literature search was based on literature titles and keywords. That could result in the addition of unwanted papers. Some legitimate AI ethics articles might be excluded.
4. The majority of literature included in this study was in English (97%). It is highly probable that some non-English literature was missed.

Despite these caveats, we believe that our bibliometric analysis remains highly valuable to the scientific and engineering community since most AI and AI Ethics research is published in English-language journals and conferences that are indexed by Scopus.

## VI. CONCLUSION

The bibliometric analysis of AI Ethics literature has pointed to a 3-phase AI Ethics development, namely incubation, making AI human-like machines, and making AI human-centric machines. AI ethicists may need to get ahead of the AI technology development and research on making AI machine-like humans, prohibit unethical development of machine-augmented non-humans, and prevent the development of malicious or malevolent Superintelligence.

## VII. ACKNOWLEDGEMENT

The work reported herein was supported by the National Science Foundation (NSF) (Award #2246920). Any opinions, findings, and conclusions or recommendations expressed in this material are those of the authors and do not necessarily reflect the views of the NSF.